\documentstyle[a4,12pt]{article}
\begin{document}

\begin{flushright} PRA-HEP-98/5
\end{flushright}

\begin{center}
{\Large {\bf Locality problem, Bell's inequalities and EPR
experiments}}\\[3mm]
{Milo\v{s} Lokaj\'{\i}\v{c}ek }\\
{Institute of Physics, AV\v{C}R, 18221 Prague 8,
Czech Republic}\\
\end{center}

\begin{abstract}
  The question has been solved whether Bell's inequalities cover all possible
  kinds of hidden-variable theories.  It has been shown that the given
  inequalities can be
 hardly derived when the changing space position of photon-pair source together
 with the microscopic space structure of measuring devices are taken into
 account; and when corresponding impact parameters (i.e., exact impact
 points) of photons in individual measuring devices (polarizers) influence
 measured values, in addition to usually considered characteristics.
\end{abstract}

\section{Introduction }

   The interpretation problem  of the
quantum-mechanical mathematical model is still open. While the
original controversy between A. Einstein and N. Bohr
\cite{ein,bohr} started from different philosophical attitudes of
these two physicists  to understanding the notion of reality it was
expected that the problem would be solved on the physical basis
after J. Bell \cite{bell} derived the known
  inequalities. These inequalities should be fulfilled by experimental data
  if a hidden-variable theory corresponded to physical reality.

    Experiments initiated by these inequalities and performed in 1972-82
(see, e.g., \cite{asp}) ended with the conclusion that Bell's inequalities
are surely violated by experimental data. And this violation has been
usually interpreted as a
decisive argument against any  theory of hidden variables and, consequently,
against the locality of microscopic objects.

    Such a conclusion might have seemed to be strongly
supported also by finding of d'Espagnat \cite{esp} that Bell's inequalities
have been practically in a conflict with Schroedinger equation itself.
 It was shown, however, already earlier by Bohm \cite{bohm} that a kind of
hidden variables has been contained already in this equation.  Therefore, the
following  question should be raised:
Are such hidden variables  excluded by experimental data, too?
Or:   Are these hidden variables covered by Bell's inequalities or not?
And/or: How to proceed to remove such a dilemma?

   Necessity of such an analysis is strongly supported also by the fact
that it is difficult to understand the nonlocality of microscopic objects
on macroscopic distances or a mechanism acting faster than light (one or
the other) as required if one accepts the standard quantum-mechanical
model and its Copenhagen interpretation.

   In the following we will attempt to give an answer  to the given
questions.  A new kind of hidden variables will be introduced and
the impossibility of deriving the usual Bell's limit under these
new conditions  will be demonstrated with the help of two standard
approaches, i.e.,  those of Bell and of Clauser and Horn (see.
e.g., \cite{clau}).   \\

\section{EPR experiment and a new kind of hidden variables}

  The contemporary EPR experiments consist in establishing  coincidence
transmissions of two photons (having the same polarization and going in
opposite directions) through two polarizers under different angles
between polarizer axes:
\begin{eqnarray}
  \alpha \hspace{3.2cm} \beta    \hspace{0.6cm}   \nonumber   \\
 {\bf |} \leftarrow --- o --- \rightarrow {\bf |}  \hspace{0.75cm}
\nonumber
\end{eqnarray}

     Coincidence probability in individual events depending on common
     hidden variables $\Lambda$  may be written  as
\begin{equation}
  p_{\alpha ,\beta}(\Lambda)  \; =
 \; p^{(1)}_{\alpha}(\Lambda) \;.\; p^{(2)}_{\beta}(\Lambda) \; .
\end{equation}
Due to incidental values of $\Lambda$  the measured values equal
\begin{equation}
    P_{\alpha,\beta} \; = \;
      \oint\! d\Lambda \; p_{\alpha,\beta}(\Lambda)
\end{equation}
where  $\oint\! d\Lambda$ represents  statistical average over all
possible quantities of $\Lambda$ set.

    The set $\Lambda$ may involve practically four different kinds of
parameters:
\begin{equation}
   \Lambda \; \equiv \; \{ \lambda, \lambda', \lambda_a, \lambda_b\}
\end{equation}
where $\lambda$ represents proper spin (or polarization) characteristic,
$\lambda' $ other hidden variables determining the state of the photon pair,
and $\lambda_a$ and $\lambda_b$ - instantaneous characteristics (including
microscopic fluctuations - see, e.g., \cite{laloe}) of individual macroscopic
measuring devices.
The last two parameters must be denoted as redundant now as
they have been practically included in probability functions when Bell
modified his original approach  \cite{bell2}. We will omit them in the
following analysis and limit our considerations to the other two.

   While the meaning of $\lambda$ is  clear we should ask what
characteristics may be represented by $\lambda'$  if they should
not be covered by Bell's inequalities (see the last paragraph of
Sec. 1). And further, whether such a characteristic may belong to a
realistic description of the microscopic world.

     Let us assume that  $\lambda'$ represents the source position of a
   photon pair and its momentum direction; i.e., three coordinates and
 two components of unit vector (energy of both the photons going in opposite
 directions being defined). These
 characteristics cannot be omitted in any realistic approach.  However, they
 make it  possible to determine the exact impact point
 of each photon in the corresponding polarizer plane, which may be
  represented by a two-dimensional vector ${\hat b}^{(j)}$ where $j=1,2$
 denotes the corresponding measuring device. Consequently, the properties
 of any individual photon in the polarizer plane are defined by
   $\lambda$ and $\hat b{(j)}$  quite independently of the setting of
 measuring devices.

    However, in a realistic picture we must admit necessarily that
 the response of a  measuring device (in a single event) will not depend
 directly on  ${\hat b}^{(j)}$ but on an exact impact point in the proper
 microscopic structure of a polarizer consisting of a periodical grid of
 atoms. And it is not possible to substitute individual results by
 statistical averages in deriving Bell's inequalities for coincidence
 measurements.

      Then of course,  the response of the measuring device in an
 individual event will depend on the proper effective impact parameter that
 will be obtained by projecting the point given by  ${\hat b}^{(j)}$
 into the polarizer two-dimensional grid structure formed by individual
 interaction centers (atoms). And it is evident that such an effective impact
 parameter will depend necessarily also on the setting $\gamma$ ($\gamma$
 representing $\alpha$ or $\beta$) due to changing correlation of the two
 given plane structures.  We will denote such an effective
 impact parameter as $b^{(j)}_\gamma$. It must be related always to an individual
 interaction center, as it is done in all collision experiments. However,
 at difference to  collision processes at higher energies
 the  orientation of impact parameters in polarizer plane structure cannot
 be neglected as the photon must interact necessarilly with more than one
 interaction center.

     Respecting, therefore, the realistic space structure of the whole
 measuring problem we must conclude that there are always three different and
     in principle independent (two-dimensional) vectors characterizing the
  situation in each measuring device in the plane perpendicular to the
  photon tracks:   $\lambda$, $b^{(1)}_\alpha$, $\alpha$; and
   $\lambda$, $b^{(2)}_\beta$, $\beta$; i.e., the spin, the effective
   impact parameter and the vector representing angle orientation of the
 given measuring devices.

   It will be shown in the following that in derivations of Bell's
   inequalities some  simplifying conditions
    have been used, which may be applied to only if the
   corresponding probabilities do not depend on the mentioned vector
   triples; e.g., if the dependence on $b^{(j)}_\gamma$ is neglected.

    While the statistical distributions of $\lambda'$ and of ${\hat b}^{(j)}$
  are fully independent of the settings of individual measuring devices
 it is not more true for the distributions of
 $b^{(j)}_\gamma$ obtained by projection into the actual polarizer
 structure. They may be  $\gamma$-dependent due
  to changing space orientations of the microscopic structures of macroscopic
  objects. We will demonstrate this new  situation now on two examples (the
approaches of Bell and  of Clauser and Horn) to a greater detail.  \\

\section{Bell's approach}
  Let us assume that the situation in each measuring device is
 characterized by the mentioned tripple of two-dimensional vectors:
$lambda$, $b^{(j)}_\gamma$.
   The  expression  for the experimentally established probabilities
 may be then written  as
\begin{equation}
 P_{\alpha,\beta}\; = \; \oint\! d\lambda \oint\!
 db^{(1)}_\alpha db^{(2)}_\beta
  \;\;  p_{\alpha,\beta}(\lambda , b^{(1)}_\alpha, b^{(2)}_\beta)
 \end{equation}
where
\begin{equation}
 p_{\alpha,\beta}(\lambda , b^{(1)}_\alpha, b^{(2)}_\beta )
   \; =  \;
   p^{(1)}_{\alpha}(\lambda ,b^{(1)}_\alpha)\; .\;
    p^{(2)}_{\beta}(\lambda ,b^{(2)}_\beta)   \;\; .
\end{equation}
   The quantity $p^{(j)}_\gamma(\lambda,b^{(j)}_\gamma)$ represents the
probability that a photon characterized by $\lambda$ and $\lambda'$
went through the corresponding polarizer  when it was set to
$\gamma$; $b^{(j)}_\gamma$ being a unique function of $\lambda'$
for a given $\gamma$.

    Bell's inequalities are then represented by the condition
\begin{equation}
     P_{\alpha,\beta}\; +\; P_{\alpha,\beta'}\; +\; P_{\alpha',\beta}
                 \;      - \;  P_{\alpha',\beta'}  \;\;  \leq \;\;  2 \;
   \label{ineq}
\end{equation}
for any values of four angles.  The inequalities (\ref{ineq}) should be
 violated in the case of the standard quantum-mechanical model.
They  have been proven with the help of different
theoretical approaches to hold for a hidden-variable theory.
  However, in none of them the influence of the mentioned effective impact
  parameters has been taken into account.

  There is a fundamental difference between the parameters
$b^{(j)}_{\gamma}$ and the previously used parameters $\lambda_a$
and $\lambda_b$. The latter ones represented time fluctuations (or
other characteristics) of individual measuring devices being fully
independent of other parameters characterizing a photon pair, while the
former ones follow from
the space structure of the whole process in each event, combining
the space orientations of the microscopic system and of measuring
devices. An important question then arises, how it is with the
derivation of Bell's inequalities for coincidence measurement in
such a case when the averaging over $\lambda'$  must be substituted
by averaging over two different effective impact parameters
$b^{(j)}_{\gamma}$. It means that while the statistical
distributions of both the parameters $\lambda$ and $\lambda'$ may
be regarded as independent of other parameters the
statistical distributions of effective impact parameters must be expected to
depend significantly on settings $\alpha$ and $\beta$; being
influenced by the corresponding orientations of space structures of
measuring devices. The $\alpha,\beta$ dependence of measured values
is then given by two different and in principle mutually
independent factors; one coming from $\lambda$ and the other coming
from $b^{(j)}_\gamma$.

   Bell's approach has started practically (see, e.g., Ref. \cite{clau}, Eqs.
(3.11-3.12)) from the equation
\begin{equation}
  P_{\alpha,\beta} \; - \;  P_{\alpha,\beta'} \; = \;
   \oint\! d\lambda\; \{ \oint\! db_\alpha db_\beta  \;  p_{\alpha,\beta}
\;-\; \oint\! db_\alpha db_{\beta'}  \;  p_{\alpha,\beta'} \}
\end{equation}
by adding to its right side the expression
\begin{equation}
 \oint\!d\lambda \; \{\oint\! db_\alpha db_\beta \;
  p^{(1)}_\alpha \; p^{(2)}_\beta \; p^{(1)}_{\alpha'} \; p^{(2)}_{\beta'}\;
          -\;  \oint\! db_\alpha db_{\beta'}  \;
 p^{(1)}_\alpha \; p^{(2)}_\beta \; p^{(1)}_{\alpha'}\; p^{(2)}_{\beta'}\}\;,
   \label{g8}
\end{equation}
 equaling zero if impact-parameter dependence is neglected. Then taking
 into account that it holds $p^{(j)}_\gamma \leq 1$ and
 inserting the values of one  instead of two  probability factors
(differently  in different terms) the inequalities (\ref{ineq})
might be derived.

   However, it is evident that the given approach cannot be applied to when
the influence of impact parameters is taken into account as the
expression (\ref{g8}) cannot be equal zero.
The expression (\ref{g8}) is well defined and may equal zero for any
$\alpha$, $\alpha'$, $\beta$ and $\beta'$   only if the statistical
distributions of all parameters $b^{j}_\gamma$ are
independent of $\gamma$;  it means if  it holds
\begin{equation}
      b^{(j)}_{\gamma} \; \equiv \; b^{(j)}_{\gamma'}
  \label{b12}
\end{equation}
for any $\gamma$ and $\gamma'$.

     Any dependence of measured values  on impact parameters  must be,
therefore, excluded if Bell's inequalities are to be derived for
experiments with two photons. Bell's inequalities would be,
however, derived if the parameter $\lambda$ were not represented by
a vector (e.g., if instead of photons two scalar particles were
emitted in opposite directions and measurement consisted in
establishing simple coincidence detection), even if Eq.
(\ref{b12}) were not fulfilled. The individual probability factors
would not depend more on mutual orientations of two vectors
$\lambda$ and $b_\gamma$ in such a case.    \\

\section{Approach of Clauser and Horn}
  The same assumption  was involved practically in the other approaches
presented in Ref. \cite{clau}, too. We
will demonstrate it yet on the approach proposed by Clauser and Horn.
 Taking into account the influence of corresponding impact parameters
one can write:
\begin{eqnarray}
P^{(1)}_{\alpha} &=& \oint\! d\lambda  \oint db^{(1)}_\alpha \;
  p^{(1)}_{\alpha}(\lambda, b^{(1)}_\alpha)  \; ,
\nonumber  \\
P^{(2)}_{\beta} &=& \oint\! d\lambda  \oint db^{(2)}_\beta  \;
      p^{(2)}_{\beta}(\lambda,b^{(2)}_\beta) \;  ,  \\
 P_{\alpha,\beta}  &=& \oint\! d\lambda \oint db^{(1)}_\alpha db^{(2)}_\beta
 \; p_{\alpha,\beta}(\lambda, b^{(1)}_\alpha, b^{(2)}_\beta )  \; .
   \nonumber
\end{eqnarray}
   It is not possible then
to obtain Eq. (3.19) of Ref. \cite{clau} simply from Eq. (3.18) as
individual members in Eq. (3.18), i.e., in the expression
\begin{eqnarray}
      p_{\alpha,\beta} -  p_{\alpha,\beta'}
       +  p_{\alpha',\beta}  +  p_{\alpha',\beta'}
        -  p^{(1)}_{\alpha}   -   p^{(2)}_{\beta}   \; ,
        \nonumber
\end{eqnarray}
 would have to be integrated and averaged over {\it different } variables
   $b^{(j)}_\gamma$ (having different statistical distributions).
 The given approach could be applied to and
Bell's inequalities would be obtained if it held again
\begin{eqnarray}
   b^{(j)}_\gamma \; \equiv\; b^{(j)}_{\gamma'}   \; .
          \label{ro12}
\end{eqnarray}

   The additional condition (\ref{b12}) or (\ref{ro12})
 relates  to one-particle states only, which is
in  agreement with  the recent results of Revzen and Mann
\cite{rev}. Any influence of internal space structure
of macroscopic measuring objects on measured results seems to be fully
neglected in such a case.    \\

\section{Concluding remark}

   The effective impact parameters  $b^{(i)}_\gamma$ must be necessarily
   included in the description of physical processes if a hidden variable
   theory is to correspond fully to physical reality. The influence of internal
   structures of macroscopic measuring devices on measured results
  cannot be then omitted. However,
   in such a case it is not more possible to derive Bell's inequalities
   for probabilities measured in EPR polarization experiments. Consequently,
   their violation by the corresponding experimental data can be  hardly
   regarded as a proof of non-locality (or inseparability) of microscopic
   objects. A more detailed analysis of the whole problem should be
   performed and  physical meaning of EPR experiments newly analyzed.

    In conclusion I should like to thank to the referees of Phys. Rev.
    Letters for valuable critical comments which helped in improving some not
    quite convincing formulations of the presented arguments.

\end{document}